\documentclass[aps,prb,twocolumn,groupedaddress,showpacs]{revtex4}

\usepackage{graphicx}
\usepackage{dcolumn}
\usepackage{bm}
\usepackage{amsmath,amssymb}
\usepackage{color}

\def\ket#1{|#1\rangle} 
\def\Vec#1{\mbox{\boldmath $#1$}}

\def\cH{{\cal H}}
\def\mtilde{\tilde{m}}

\begin{document}


\title{Translational symmetry broken magnetization plateau of the $S=1$ antiferromagnetic Heisenberg chain with competing anisotropies}



\author{T\^oru Sakai,$^{1,2}$ Kiyomi Okamoto$^{1}$, Kouichi Okunishi$^{3}$, Masaru Hashimoto$^1$, Tomoki Houda$^1$, Rito Furuchi$^1$ and Hiroki Nakano$^1$ }
\affiliation{$^1${Graduate School of Science, University of Hyogo, Kouto 3-2-1, Kamigori, Ako-gun, Hyogo 678-1297 Japan}\\
$^2${National Institute for Quantum Science and Technology (QST) SPring-8, Kouto 1-1-1, Sayo, Sayo-gun, Hyogo 679-5148 Japan}\\
$^3${Department of Physics, Niigata University, Niigata 950-2181, Japan} 
}


\date{\today}

\begin{abstract}

We investigate the $S=1$ antiferromagnetic quantum spin chain with the 
exchange and single-ion anisotropies in a magnetic field, using the 
numerical exact diagonalization of finite-size clusters, the 
level spectroscopy analysis, and the density matrix renormalization group 
(DMRG) methods. 
It is found that a translational symmetry broken magnetization plateau possibly appears at the half of 
the saturation magnetization, when the anisotropies compete with each other. 
The level spectroscopy analysis gives the phase diagram at half the saturation magnetization. 
The DMRG calculation presents the magnetization curves for some typical 
parameters and clarifies the spin structure in the plateau phase. 

\end{abstract}


\maketitle


\section{Introduction}

One-dimensional quantum spin systems have been attracting increasing attention 
both experimentally and theoretically in recent years.\cite{TLL}
There have been found various kinds of phenomena which was originated from the strong
spin-spin interactions as well as the strong quantum fluctuations due to one dimension.
Among these phenomena,
the magnetization plateau is one of most interesting phenomena 
because it is a macroscopic quantized phenomenon with a topological background in many body spin systems.
In the quantum spin chains,
based on the Lieb-Schultz-Mattis theorem\cite{LSM}, 
the rigorous necessary condition for the appearance of the plateau 
was derived as the form\cite{oshikawa}
\begin{eqnarray}
Q(S-\mtilde)={\rm integer},
\label{condition}
\end{eqnarray}
where $S$ and $\mtilde$ are the total spin and the magnetization per unit cell, and 
$Q$ is the periodicity of the ground state measured by the unit cell. 
The magnetization plateau for $Q\ge 2$ should be accompanied by 
the spontaneous translational symmetry breaking. 
The simple magnetization plateau for $Q=1$ has been theoretically predicted 
or experimentally observed in the following systems; 
the $S=3/2$ and $S=2$ anisotropic antiferromagnetic chains\cite{sakai1,kitazawa}, 
the $S=1/2$ distorted diamond chain\cite{honecker1,okamoto1,kikuchi,gu-su,honecker2,ananikian,morita,ueno,filho}, 
the $S=1/2$ trimerized chain\cite{hida,okamoto-ssc,oka-kita,gong2,liu2}, 
the $S=1/2$ tetramerized chain\cite{gong,mahdavifar,jiang,liu3}, 
the $S=1/2$ two-leg ladder\cite{sugimoto3,sugimoto1,sugimoto2,sasaki,rahaman},
the $S=1/2$ three-leg spin ladder and tube\cite{cabra,okamoto-tube,li,alecio,farchakh},
the $S=1/2$ and $S=1$ skewed systems\cite{yin,dey},
the mixed spin chain\cite{yamamoto,sakai2,tonegawa,tenorio,liu,karlova,yamaguchi}, 
the $p$-leg ladder\cite{cabra3},
the polymerized chain\cite{cabra2,chen}
etc. 

For the $S=1$ chain case,
when the unit cell is composed of one $S=1$ spin,
the magnetization plateau at half of the saturation is impossible with $Q=1$
because Eq.(\ref{condition}) cannot be satisfied with $S=1$ and $\mtilde=1/2$.
Thus the unit cell should be composed of two (more generally even number) $S=1$ spins (namely dimerization)
for the realization of this half plateau.
In this case the parameter set $S=2$ and $\mtilde=1$ satisfies Eq.(\ref{condition}) with $Q=1$.
In fact,
the half magnetization plateaus were experimentally observed in several $S=1$ chain materials with the dimerization
\cite{narumi,maximova}.
A phase diagram on the plane of the dimerization parameter versus the magnetization was
numerically obtained by Yan et al.\cite{yan}

The translational symmetry broken plateau for $Q\ge 2$ also has been 
revealed to appear in the following systems; 
the $S=1/2$ frustrated bond-alternating chain\cite{totsuka}, 
the $S=1/2$ zigzag chain\cite{okunishi1,okunishi2,metavitsiadis}
the $S=1$ frustrated chain\cite{nakano}, 
the $S=1/2$ frustrated spin ladder\cite{okazaki1,okazaki2,nakasu,sakai3,sugimoto3,sugimoto1,sugimoto2,sasaki}, 
the $S=1$ frustrated spin ladder\cite{okamoto2,okamoto3,michaud,kohshiro}, etc. 
In most cases, the mechanism of the $Q\ge 2$ plateau has been based on the frustration. 
Recently the numerical diagonalization study on the $S=2$ antiferromagnetic chain 
indicated that the competing anisotropies possibly yields the 
$Q=2$ plateau at half the saturation magnetization\cite{yamada}, 
as well as the $Q=1$ plateau.
Thus the competing anisotropies are expected to give rise to the $Q = 2$ plateau, even without frustration. 

However, the half magnetization plateau of $S=1$ chain without dimerization 
(namely, $Q=2$, $S=1$, $\mtilde=1/2$) has not been observed so far\cite{maximova} as far as we know,
Thus we think that it is important to clarify the condition for the realization
of the half plateau in the $S=1$ spin chains with $Q=2$, $S=1$, $\mtilde=1/2$.

Considering the above situation,
in this paper we investigate the $S=1$ antiferromagnetic chain with the 
$XXZ$-coupling and single-ion anisotropies competing with each other, 
and clarify the condition for the $Q=2$ plateau at half the saturation 
magnetization. 
This may give the reason why such a plateau has not been experimentally observed,
as well as provide a guide for finding or synthesizing the materials showing such a plateau.
Using the numerical diagonalization of finite-size clusters and the level spectroscopy analysis, 
the phase diagram at half the saturation magnetization is presented. In addition the density matrix 
renormalization group (DMRG) calculation indicates that the $Q=2$ plateau actually appears on the 
magnetization curve. 
We also show the phase diagram of the magnetization process.

\section{Model}

We investigate the magnetization process 
of the $S=1$ antiferromagnetic Heisenberg chain with the 
exchange and single-ion anisotropies, denoted by $\lambda$ and $D$, 
respectively. 
The Hamiltonian is given by 
\begin{eqnarray}
\label{ham}
&{\cal H}&={\cal H}_0+{\cal H}_Z, \\
&{\cal H}_0& = \sum _{j=1}^L \left[ S_j^xS_{j+1}^x + S_j^yS_{j+1}^y 
 + \lambda S_j^zS_{j+1}^z \right] \nonumber \\
  &&+D\sum_{j=1}^L (S_j^z)^2, \\
&{\cal H}_Z& =-H\sum _{j=1}^L S_j^z.
\end{eqnarray}
The exchange interaction constant is set to be unity as the unit of energy.
For $L$-site systems, 
the lowest energy of ${\cal H}_0$ in the subspace where 
$\sum _j S_j^z=M$, is denoted as $E(L,M)$. 
The reduced magnetization $m$ is defined as $m=M/M_{\rm s}$, 
where $M_{\rm s}$ denotes the saturation of the magnetization, 
namely $M_{\rm s}=L$. 
$E(L,M)$ is calculated by the Lanczos algorithm under the 
periodic boundary condition ($ \Vec{S}_{L+1}=\Vec{S}_1$). 
We consider the case when $\lambda$ is Ising-like and 
$D$ is $XY$-like,  
namely, $\lambda > 1$ and $D >0$. 
Thus easy-axis $\lambda$ and easy-plane $D$ are competing with each other. 
If the magnetization plateau appears at $m=1/2$, 
the translational symmetry should be spontaneously broken 
and the two-fold degeneracy of the ground state should occur, 
namely $Q=2$.

\section{Phase diagram at $m=1/2$}

In this section using the numerical diagonalization for finite-size clusters, 
the phenomenological renormalization group and the level spectroscopy analyses, 
we show that the magnetization plateau appears at $m=1/2$ for sufficiently 
large $\lambda$ and $D$, and present the phase diagram at $m=1/2$. 

\subsection{Phenomenological Renormalization Group}

\begin{figure}
\includegraphics[width=0.85\linewidth,angle=0]{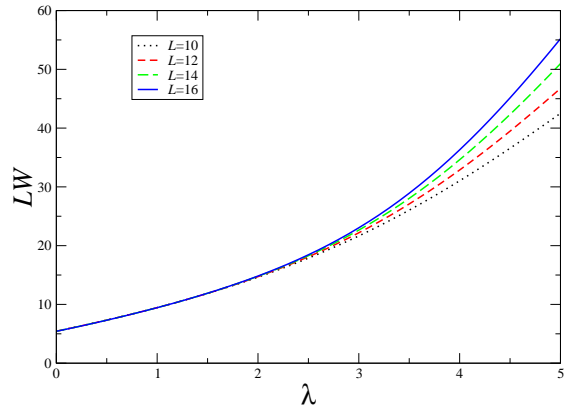}%
\caption{\label{lw} 
Scaled plateau width $LW$ plotted versus $\lambda$ 
for $L=$10, 12, 14 and 16 in the case of $D=5.0$. 
}
\end{figure}

In order to confirm that the magnetization plateau really appears 
at $m=1/2$, we apply the phenomenological renormalization group\cite{PRG}
for the plateau width $W$ defined as the form
\begin{eqnarray}
W=E(L,M-1)+E(L,M+1)-2E(L,M), 
\label{w}
\end{eqnarray}
where $M=L/2$. 
Since $W$ should be proportional to $1/L$ in the no-plateau case, 
the scaled width $LW$ would be independent of the system size $L$, 
while $W$ would increase with $L$ in the presence of plateau. 
Let us set $D=5.0$ as an example.
With fixed $D=5.0$, $LW$ calculated for $L=$10, 12, 14 and 16 are 
plotted versus $\lambda$ in Fig.~\ref{lw}. 
It indicates that the plateau obviously appears for sufficiently 
large $\lambda$. However, it is difficult to determine the 
precise phase boundary with this method. 

\begin{figure}
\includegraphics[width=0.85\linewidth,angle=0]{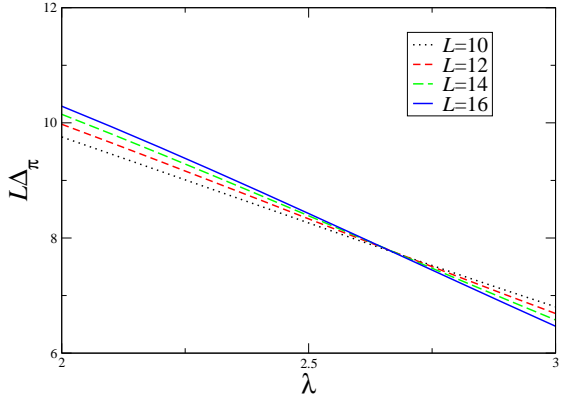}%
\caption{\label{pi} 
Scaled gap $L\Delta_{\pi}$ plotted versus $\lambda$ 
for $L=$10, 12, 14 and 16 in the case of $D=5.0$. 
}
\end{figure}

\begin{figure}
\bigskip
\includegraphics[width=0.85\linewidth,angle=0]{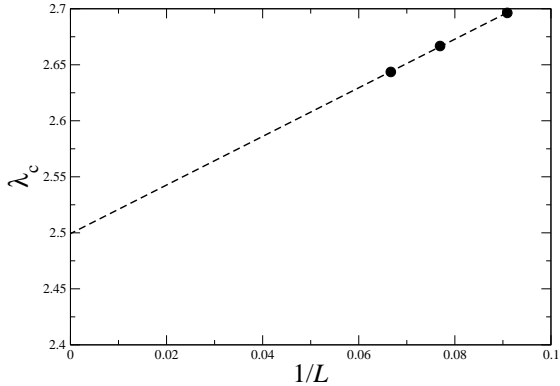}%
\caption{\label{gaisoprg} 
Size-dependent fixed points $\lambda_{\rm c}(L)$ obtained by
the phenomenological renormalization group method
for $L=$11, 13 and 15 are plotted versus $1/L$ 
for $D=5.0$. 
The phase boundary in the thermodynamic limit is estimated as 
$\lambda_{\rm c}=2.50 \pm 0.01$. 
}
\end{figure}

Next, we apply the phenomenological renormalization group analysis \cite{PRG}
for the excitation gap with the momentum $k=\pi$ in the subspace 
$m=1/2$, defined as $\Delta _{\pi}$. 
The size-dependent fixed point $\lambda_{\rm c}(L+1)$ is determined by 
the equation
\begin{eqnarray}
L\Delta _{\pi}(L,\lambda)=(L+2)\Delta_{\pi}(L+2,\lambda).
\label{prgpi}
\end{eqnarray}
The scaled gaps $L\Delta_{\pi}$ for $D=5.0$ are plotted versus $\lambda$ for 
$L=$10, 12, 14 and 16 in Fig.~\ref{pi}. 
The size-dependent fixed points $\lambda_{\rm c}(L)$ for $L=$11, 13 and 15 are plotted versus $1/L$ 
for $D=5.0$ in Fig.~\ref{gaisoprg}. 
The phase boundary in the thermodynamic limit is estimated as 
$\lambda_{\rm c}=2.50 \pm 0.01$. 
We repeat this procedure for various fixed $D$ or for fixed $\lambda$ to estimate the phase boundary.
Actually, the phase boundary for $D\ge 3.0$ was obtained by fixed $D$ method, 
while that for $\lambda \ge 3.5$ estimated from the $\lambda$ method. 
The present result suggests that the translational symmetry is spontaneously broken 
and the ground state has a two-fold degeneracy in the plateau phase. 
The N\'eel order like $|\cdots 101010 \cdots \rangle$ is expected to be realized. 
Thus we call this plateau "N\'eel plateau".

\subsection{Level spectroscopy}

\begin{figure}
\includegraphics[width=0.85\linewidth,angle=0]{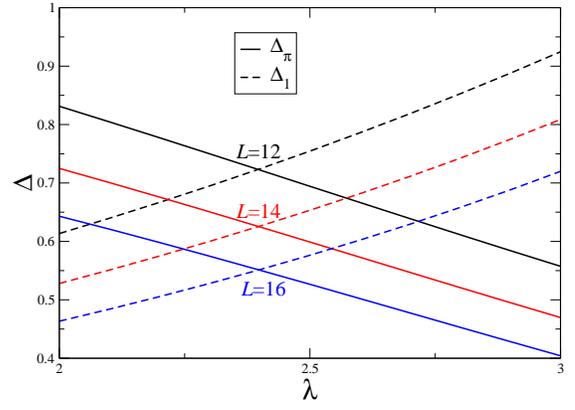}%
\caption{\label{LSd50} 
$\Delta_1$ and $\Delta_{\pi}$ for $D=5.0$ are plotted versus $\lambda$ for $L=$12, 14 and 16. 
}
\end{figure}

\begin{figure}
\bigskip
\includegraphics[width=0.85\linewidth,angle=0]{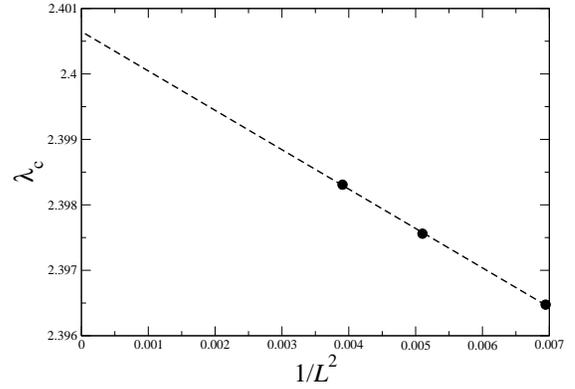}%
\caption{\label{gaisod50} 
The extrapolation of $\lambda_{\rm c}$ to the thermodynamic limit, assuming the size correction if proportional to $1/L^2$,
gives $\lambda_{\rm c}=2.401 \pm 0.001$.
}
\end{figure}

One of more precise methods to determine the phase boundary is 
the level spectroscopy analysis.\cite{oka-nom,nom-oka}
Based on this method, comparing the single magnon excitation gap $\Delta_1 \equiv W/2$ and $\Delta_{\pi}$, 
the gap $\Delta_1$ is smaller in the no-plateau phase, while $\Delta_{\pi}$ is smaller in the plateau phase. 
Thus $\Delta_1=\Delta_{\pi}$ gives the size-dependent phase boundary. 
$\Delta_1$ and $\Delta_{\pi}$ for $D=5.0$ are plotted versus $\lambda$ for $L=$12, 14 and 16 
in Fig. \ref{LSd50}. 
It indicates $L$ dependence is quite small and the size correction is predicted to be proportional to $1/L^2$. 
The extrapolation of $\lambda_{\rm c}$ to the thermodynamic limit gives $\lambda_{\rm c}=2.401 \pm 0.001$, as shown in Fig.~\ref{gaisod50}. 
Although there is a small discrepancy of the extrapolated phase boundary
between the phenomenological renormalization and the 
level spectroscopy because of some finite-size effect, 
the latter method is expected to be more precise, 
because it is based on the essential nature of the Berezinskii-Kosterlitz-Thouless
transition.\cite{berezinskii,KT,oka-nom,nom-oka,TLL,jose}
Namely the lowest order contributions of the logarithmic size corrections are cancelled out
with each other in the level spectroscopy method.\cite{oka-nom,nom-oka}

\subsection{Magnetization jump}
\begin{figure}
\includegraphics[width=0.85\linewidth,angle=0]{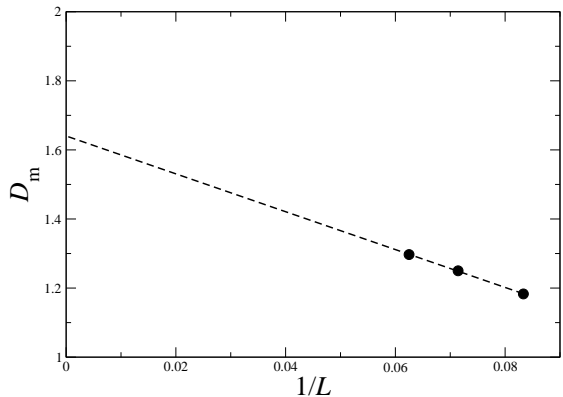}%
\caption{\label{extradm80} 
The boundary of the missing region $D_{\rm m}$ for $\lambda =8.0$ is plotted versus $1/L$. 
As assuming the size correction proportional to $1/L$, $D_{\rm m}$ in the infinite length limit is 
estimated as $D_{\rm m} = 1.64 \pm 0.01$. 
}
\end{figure}

Apart from the no-plateau and the magnetization plateau phases, there is a parameter region 
where the $m = 1/2$ magnetization is not realized due to the magnetization jump.
like the spin flop transition. 
A typical case for the "missing" can be seen in the magnetization curve of
$\lambda=8.0$ and $D=0.0$ of Fig.\ref{ising}.
There is a magnetization jump from about $m=0.04$ to $m=0.55$,
which means that the $m=1/2$ situation is not realized in this curve.
If the $m = 1/2$ magnetization is included in the magnetization jump, 
we call that the system is in the missing region. 
The boundary of the missing region $D_{\rm m}$ for $\lambda =8.0$ is plotted versus $1/L$ in Fig.~\ref{extradm80}. 
Assuming the size correction proportional to $1/L$, $D_{\rm m}$ in the infinite length limit is estimated as $D_{\rm m}=1.64 \pm 0.01$. 
The boundary of the missing region is determined by this method.

\subsection{Phase diagram}

\begin{figure}
\bigskip
\includegraphics[width=0.85\linewidth,angle=0]{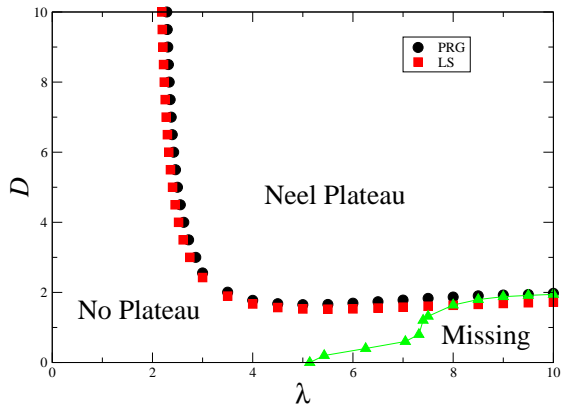}%
\caption{\label{phase} 
Phase diagram at $m=1/2$ of the present model. 
There are the no-plateau, N\'eel plateau phases and the missing region which is surrounded by green triangles. 
}
\end{figure}

Here we present the phase diagram at half the saturation magnetization with respect to the 
easy-axis coupling anisotropy $\lambda$ and the easy-plane single-ion one $D$ in Fig.~\ref{phase}. 
It consists of the no-plateau, N\'eel plateau phases and the missing region which is surrounded by green triangles. 
In the N\'eel plateau phase the translational symmetry is spontaneously broken and $Q=2$ is realized. 

\section{Magnetization curves}
\begin{figure}
\bigskip
\includegraphics[width=0.70\linewidth,angle=0]{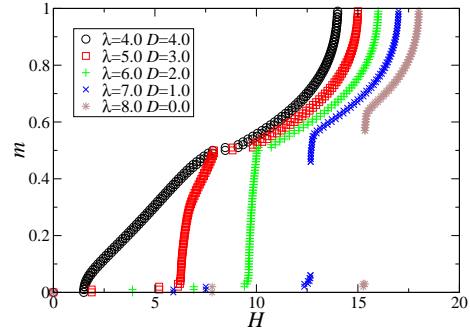}%
\caption{\label{ising} 
The magnetization curves calculated by DMRG  for
$(\lambda, D)=(4.0, 4.0),~(5.0, 3.0),~(6.0, 2.0),~(7.0, 1.0)$ and $(8.0, 0.0)$ 
are shown 
by black circles, red squares, green pluses, blue crosses and brown stars, respectively. 
}
\end{figure}

In order to confirm that the 1/2 magnetization plateau actually appears, 
we performed the DMRG calculation with $L=100$ to obtain the magnetization curves 
in the ground state. 
The calculated magnetization curves for
($\lambda, D)=(4.0, 4.0),~(5.0, 3.0),~(6.0, 2.0),~(7.0, 1.0)$ and $(8.0, 0.0)$
are shown in Fig.~\ref{ising},
by black circles, red squares, green pluses, blue crosses and brown stars, respectively.
The curves for (5.0, 3.0) and (6.0, 2.0) in the plateau phase obviously exhibit the 1/2 magnetization plateau. 
On the curve for $(8.0, 0.0)$  the $m=1/2$ state is skipped due to the magnetization jump. 
For the case of $(7.0,1.0)$ the $m=1/2$ state is realized,
although there is a magnetization jump.

\begin{figure}
\includegraphics[width=0.70\linewidth,angle=0]{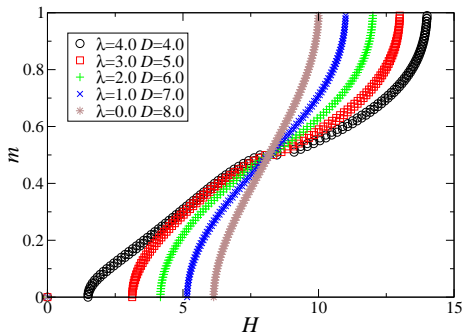}%
\caption{\label{largeD} 
The magnetization curves calculated by DMRG  for
$(\lambda, D)=(4.0, 4.0),~(3.0, 5.0),~(2.0, 6.0),~(1.0, 7.0)$ and $(0.0, 8.0)$ 
are shown by
by black circles, red squares, green pluses, blue crosses and brown stars, respectively. 
}
\end{figure}

The magnetization curves by DMRG for
$(\lambda, D)=(4.0, 4.0),~(3.0, 5.0),~(2.0, 6.0),~(1.0, 7.0)$ and $(0.0, 8.0)$ 
are also shown in Fig.~\ref{largeD}, 
by black circles, red squares, green pluses, blue crosses and brown stars, respectively. 
The curves for  (0.0, 8.0), (1.0, 7.0) and (2.0, 6.0) in the no-plateau phase have no plateau, 
while the ones for (3.0, 5.0) and (4.0, 4.0) in the plateau phase exhibit the $1/2$ plateau. 
These magnetization curves are all consistent with the phase diagram in Fig.~\ref{phase}. 

The saturation field $H_{\rm s}$ can be calculated from the energy difference 
between the energy of the ferromagnetic state and that of the 1-spin-down state
of the Hamiltonian (\ref{ham}).
A simple calculation leads to
\begin{equation}
   H_{\rm s} = 2\lambda + D + 2.
\end{equation}
All the magnetization curves of Figs.\ref{ising} and \ref{largeD} were calculated under the condition
$\lambda + D = 8$,
which leads to
\begin{equation}
   H_{\rm s} = \lambda + 10.
\end{equation}
This well explains all of $H_{\rm s}$ in Figs.~\ref{ising} and \ref{largeD}.

\section{Spin structure}

\begin{figure}
\includegraphics[width=0.65\linewidth,angle=0]{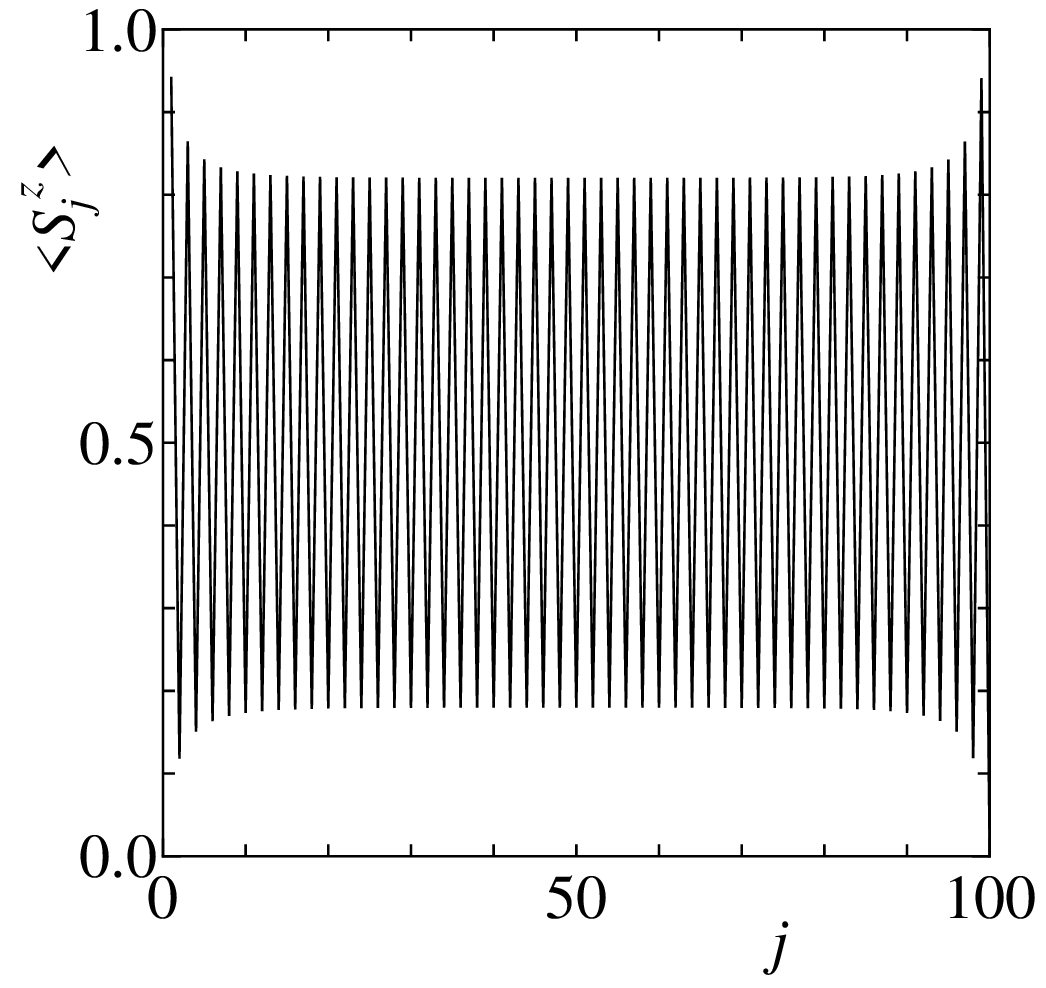}%
\caption{\label{site-mag}
Site magnetization $\langle S_j^z \rangle$ for $(\lambda,D) = (4.0,4.0)$
in the N\'eel plateau phase by DMRG with $L=100$.
We can see the N\'eel-type structure $|\cdots 101010 \cdots \rangle$.
}
\end{figure}

In order to investigate the spin structure at the 1/2 magnetization plateau, 
we calculated the magnetization at each site by DMRG. 
The site magnetization $\langle S_j^z \rangle$ at $m=1/2$ for ($\lambda, D)=(4.0, 4.0)$ in the plateau phase 
is shown in Fig.~\ref{site-mag}. 
It indicates that the translational symmetry is spontaneously broken and the periodicity $Q=2$ is 
realized. 
It is consistent with the physical picture of the N\'eel plateau. 

\section{Effective Theory }

Let us start with the isolated spin limit to construct an effective theory.
For the case of $m=1/2$,
the $\ket{S^z=0}$ state and the $\ket{S^z=1}$ state have the same energies,
which are lower than the energy of the $\ket{S^z=-1}$ state by $2D$.
We can construct an effective theory by picking up only the $\ket{S^z=0}$ state and the $\ket{S^z=1}$ state
when $D$ is sufficiently larger than the interactions, namely
\begin{equation}
    D \gg \lambda.
    \label{eq:eff-cond}
\end{equation}
We introduce the pseudo-spin operator $\Vec{T}$ with $T=1/2$,
where $\ket{T^z=1/2}$ and $\ket{T^z=-1/2}$ represent $\ket{S^z=1}$ and $\ket{S^z=0}$, respectively.
In this restricted basis,
we see
\begin{equation}
   S^z = T^z + {1 \over 2},~~~~S^\pm = \sqrt{2}T^\pm.
\end{equation}
Therefore we obtain the effective Hamiltonian as
\begin{eqnarray}
   \cH_{\rm eff}
   &=& \sum_{j=1}^L \left\{ 2(T_j^x T_{j+1}^x + T_j^y T_{j+1}^y) + \lambda T_j^z T_{j+1}^z \right\} \nonumber \\
   &&  + (\lambda + D - H)\sum_{j=1}^L T_j^z + L{\lambda +D + H \over 4}.
   \label{eq:Heff} 
\end{eqnarray}
The condition $\sum_j T_j^z = 0$ corresponds to $m=1/2$ of the original model.
From the exact solution,\cite{TLL}
the ground-state of $\cH{\rm eff}$ for $\sum_j T_j^z = 0$ is either the Tomonaga-Luttinger liquid state\cite{TLL}
(no plateau of the original model) or the N\'eel state (plateau with the N\'eel mechanism of the original model)
according as $\lambda \le 2$ or $\lambda > 2$.
We note that there is a factor 2 in front of $T_j^x T_{j+1}^x + T_j^y T_{j+1}^y$ in Eq.(\ref{eq:Heff}).
Thus the behavior of the boundary between the plateau and no plateau phase $\lambda \to 2$ as $D \to \infty$
in Fig.~\ref{phase} is well explained.
The magnetic field $H_{1/2}$ corresponding to $m=1/2$ can be obtained from the condition
that the effective field for the $T$-system is zero, namely $\lambda + D - H_{1/2} = 0$,
resulting in
\begin{equation}
   H_{1/2} = \lambda + D. 
\end{equation}
For the magnetization curves of Figs.~\ref{ising} and \ref{largeD},
we set $\lambda + D = 8$.
Then DMRG results $H_{1/2} \simeq  8$ for all the curves of Fig.~\ref{largeD} are
also well explained by this effective theory.
For the magnetization curves in Fig.~\ref{ising},
this effective theory does not hold because Eq.(\ref{eq:eff-cond}) is not satisfied.

In the phase diagram Fig.\ref{phase},
we see that two features in the $\lambda \to \infty$ limit.
One is that (a) the plateau-no plateau line and the missing boundary line
are going to merge, 
and the other is that (b) the critical value of $D$ tends to $D_{\rm c} \simeq 2$.
Liu et al.\cite{liu-ssc} investigated the phase diagram of the $S=1$ Ising chain
\begin{equation}
   {\cal H}
   = \sum_{j=1}^L S_j^z S_{j+1}^z + D_0 \sum_{j=1}^L (S_j^z)^2 - H \sum_{j=1}^L S_j^z.
   \label{eq:liu}
\end{equation}
to obtain the phase diagram on the $D_0-H$ plane.
The feature (a) is consistent with the phase diagram of Liu et al.,
although the feature (b) cannot be explained by it
since the transverse coupling is not included their Hamiltonian (\ref{eq:liu}).



\section{Phase diagram of magnetization process}

\begin{figure}
\includegraphics[width=0.85\linewidth,angle=0]{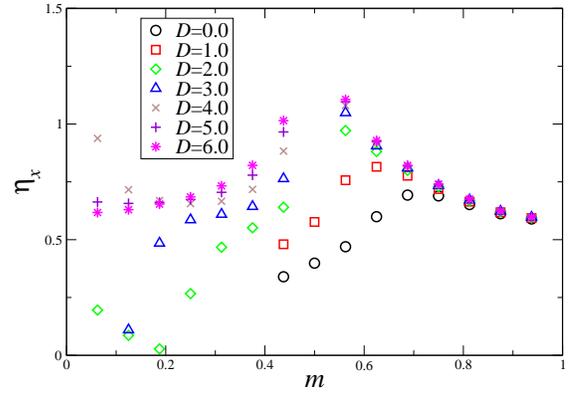}%
\caption{\label{eta} 
Exponent $\eta_x$ estimated by the numerical exact diagonalization of the 16-spin system for $\lambda=4.0$ 
plotted versus $m$ for $D=$0.0 (black circles), 1.0 (red squares), 2.0 (green diamonds), 3.0 (blue triangles), 
4.0 (brown crosses), 5.0 (violet pluses) and 6.0 (pink stars). 
}
\end{figure}

\begin{figure}
\includegraphics[width=0.85\linewidth,angle=0]{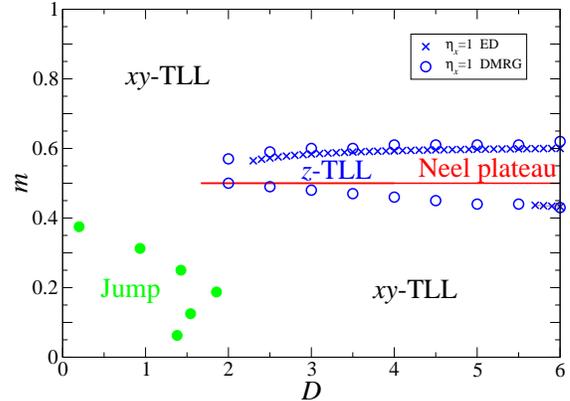}%
\caption{\label{dmphase} 
Phase diagram with respect to the anisotropy $D$ and the magnetization $m$.  
The crossover lines between the incommensurate parallel spin correlation dominant Tomonaga-Luttinger liquid 
($z$TLL) phase and the N\'eel-like perpendicular correlation dominant one ($xy$TLL) are estimated by 
the numerical exact diagonalization (ED) (blue crosses) and the DMRG (blue circles). 
The phase boundary of the missing region because of the magnetization jump is estimated by the numerical 
exact diagonalization for $L=16$ (green circles). 
The N\'eel plateau is realized just on the red line. 
}
\end{figure}

In order to consider some realistic experiments, it would be useful to obtain the phase diagram of the 
magnetization process summarizing the spin structure. 
In the gapless phase of the magnetization process, the system is expected to be in the 
Tomonaga-Luttinger liquid phase. It is characterized by the power-law decay of the spin 
correlation functions which have the asymptotic forms 
\begin{eqnarray}
&&\langle S_0^z S_r^z \rangle -m^2 \sim \cos(2k_{\rm F}r)r^{-{\eta}_z} , \\ 
&&\langle S_0^x S_r^x \rangle \sim (-1)^r r^{-\eta_x} 
\label{correlation}
\end{eqnarray}
in the infinite $r$ limit. 
$2k_{\rm F}$ is $\pi(1-m)$ in the present model. 
The first equation corresponds to the SDW spin correlation parallel to the external field 
and the second one corresponds to the N\'ee-like spin correlation perpendicular to the external field. 
The smaller exponent between $\eta_z$ and $\eta_x$ determines the dominant spin correlation. 
In the conventional magnetization process the canted N\'eel-like spin correlation is dominant, 
namely $\eta_x < \eta_z$. 
However, in some frustrated systems the magnetization region where $\eta_z < \eta_x$ is realized  appears 
and the incommensurate spin correlation parallel to the external field is dominant there.\cite{maeshima}
Then we consider the possibility of a similar interesting behavior in the present model. 
According to the conformal field theory these exponents can be estimated by the forms\cite{cardy}
\begin{eqnarray}
&&\eta_x={{E(L,M+1)+E(L,M-1)-2E(L,M)}\over{E_{k_1}(L,M)-E(L,M)}}, \\
&&\eta_z=2{{E_{2k_F}(L,M)-E(L,M)}\over{E_{k_1}(L,M)-E(L,M)}},
\label{exponent}
\end{eqnarray}
for each magnetization $M$, where $k_1$ is defined as $k_1=L/2\pi$. 
Since the relation $\eta_x \eta_z =1$ is satisfied in the Tomonaga-Luttinger liquid phase, 
we have only to calculate one of these two exponents to determine the dominant spin 
correlation. 
We estimate the exponent $\eta_x$ here, because the calculation of $\eta_z$ meets 
the larger finite-size correlation due to the incommensurate correlation expressed by the cosine factor in
Eq. (\ref{correlation}). 
The estimated exponent $\eta_x$ by the numerical exact diagonalization for $L=16$ and $\lambda =4.0$ is 
plotted versus the magnetization $m$ for several values of $D$ in Fig. \ref{eta}. 
In the case of $D \ge 3.0$, the magnetization region where $\eta_x$ is larger than 1 appears around $m\sim 1/2$. 
It indicates that the $z$ component dominant Tomonaga-Luttinger liquid phase takes place. 
Using the numerical exact diagonalization for $L=16$, $\eta_x$ can be calculated for $M=1, 2, \cdots, 15$. 
Then we estimate the crossover line $\eta_x=1$,
interpolating linearly the calculated values of $\eta_x$ at $M$ and $M+1$ 
between which $\eta_x=1$ would occur. 
In addition we estimate the critical point $D_c$ where the magnetization jump begins at each $M$ 
using the numerical exact diagonalization for $L=16$. 
The estimated crossover line between the $z$ component dominant Tomonaga-Luttinger liquid ($z$TLL) phase 
and the $xy$ component dominant one ($xy$TLL), and the critical line of the magnetization jump are 
shown in the $D$ and magnetization phase diagram for $\lambda=4.0$ in Fig. \ref{dmphase}. 
In order to confirm whether the crossover line really exists even in the thermodynamic limit, 
we also calculate $\langle S_j^xS_{j+r}^x\rangle$ for the central region of an $L=100$ chain with DMRG,
and then estimate the exponent of its power-law decay for $r = 1 \sim 30$. 
These crossover lines estimated by the numerical exact diagonalization and by the DMRG 
are shown as blue crosses and blue circles, respectively in Fig. \ref{dmphase}. 
They are consistent with each other and it suggests that the $z$TLL phase is realized 
even in the infinite length limit. 
In conclusion, it is found that the present competing anisotropies give rise to the 1/2 translational symmetry broken 
magnetization plateau and the incommensurate parallel spin correlation dominant Tomonaga-Luttinger liquid ($z$TLL) 
phase around the plateau. 
Even for different $\lambda$, qualitatively similar phase diagrams would be obtained. 


\section{Summary}
The magnetization process of the $S=1$ antiferromagnetic chain with the easy-axis coupling anisotropy 
and the easy-plane single-ion anisotropy is investigated using the numerical diagonalization for 
finite-size clusters and the DMRG calculations. 
It is found that the translational symmetry broken magnetization plateau 
appears at half the saturation magnetization for very large anisotropies (both of $\lambda$ and $D$). 
This explains the reason why this plateau has not yet been found in the $S=1$ chain compounds
without the dimerization.\cite{maximova}
Several typical magnetization curves are also presented. 
Then, the effective theory constructed for $D \gg \lambda$ well explains the numerical results in Fig. \ref{phase}.
Nevertheless, an effective theory for the $D \ll \lambda$ case and the magnetization jump is a future problem.
In addition, it is shown that the unconventional incommensurate parallel spin correlation dominant ($\eta_x > \eta_z$)
Tomonaga-Luttinger liquid phase also appears around the 1/2 plateau as in Fig. \ref{dmphase}. 
This situation is very natural because the condition for the realization\cite{TLL} of the N\'eel state ($|\cdots 101010 \cdots \rangle$)
is both of $\eta_x > \eta_z$ and the commensurability which is satisfied only at $m=1/2$.

In the previous work,\cite{yamada}
we investigated the half-plateau problem of a similar model but with $S=2$
to obtain the phase diagram
which was much richer than Fig. \ref{phase} of this paper.
In fact, the Haldane plateau phase and the large-$D$ plateau phase appeared in the $S=2$ case.
This is because the half plateau is possible without the spontaneous breaking of the
translational symmetry for the $S=2$ case.
Namely the condition (\ref{condition}) can be satisfied by $Q=1$, $S=2$, $\tilde m =1$
(note that $\tilde m =1$ for the half plateau of the $S=2$ chain).

From the experimental point of view, one can usually expect a weak
interchain interaction, which may induce the spin order corresponding
to the most dominant correlation at a low but finite temperature.
The phase diagram of Fig. \ref{dmphase} suggests that the incommensurate-SDW order
associated with the $z$TLL can be realized around the $m=1/2$ plateau
in the broad parameter region.
Thus, such an enhancement of the SDW order could be a signature of
the $m=1/2$ plateau due to the N\'eel-type mechanism, even if the width of the plateau is very narrow. 
We believe that the phase diagrams of Figs. \ref{phase} and \ref{dmphase} will be a powerful guideline
for searching or synthesizing quasi-one-dimensional materials with $S=1$
which exhibit the half plateau without the dimerization.

\begin{acknowledgments}

This work was partly supported by KAKENHI, 
Grant Numbers JP16K05419, JP20K03866, JP16H01080 (J-Physics), 
JP18H04330 (J-Physics), JP20H05274 and 23K11125 from JSPS of Japan,
and also by a Grant-in-Aid for Transformative Research Areas "The Natural Laws of Extreme Universe ---A New Paradigm
for Spacetime and Matter from Quantum Information" (KAKENHI Grant No. JP21H05191) from MEXT of Japan.
A part of the computations were performed using
facilities of the Supercomputer Center,
Institute for Solid State Physics, University of Tokyo,
and the Computer Room, Yukawa Institute for Theoretical Physics,
Kyoto University.
We used the computational resources of the supercomputer 
Fugaku provided by the RIKEN through the HPCI System 
Research projects (Project ID: hp200173, hp210068, hp210127, 
hp210201, hp220043, and hp230114). 
\end{acknowledgments}


\begin{thebibliography}{99}



%

\bibitem{TLL}
for a review,
T. Giamarchi, {\it Quantum Physics in One Dimension}, (Clarendon Press, Oxford, 2003).

\bibitem{LSM}
E. H. Lieb, T. Schultz and D. J. Mattis, Ann. Phys. (N.Y.) {\bf 16}, 
407 (1961). 

\bibitem{oshikawa}
M. Oshikawa, M. Yamanaka and I. Affleck, Phys. Rev. Lett. {\bf 78}, 
1984 (1997). 


\bibitem{sakai1}
T. Sakai and M. Takahashi, Phys. Rev. B {\bf 42}, 4537 (1990). 

\bibitem{kitazawa}
A. Kitazawa and K. Okamoto, Phys. Rev. B {\bf 62}, 940 (2000). 

\bibitem{honecker1}
A. Honecker and A. L\"auchli, Phys. Rev. B {\bf 63}, 174407 (2001).


\bibitem{okamoto1}
K. Okamoto, T. Tonegawa and M. Kaburagi, J. Phys.: Condens. Matter {\bf 15}, 
5979 (2003). 

\bibitem{kikuchi}
H. Kikuchi, Y. Fujii, M. Chiba, S. Mitsudo, T. Idehara, T. Tonegawa, K. Okamoto, T. Sakai, T. Kuwai, and H. Ohta, 
Phys. Rev. Lett. {\bf 94}, 227201 (2005). 

\bibitem{gu-su}
B. Gu and G. Su, Phys. Rev. B {\bf 75}, 174437 (2007).

\bibitem{honecker2}
A. Honecker, S. Hu, R. Peters, and J. Richter,
J. Phys.: Condens. Matter {\bf 23},164211 (2011).

\bibitem{ananikian}
N. S. Ananikian, J. Stre\v{c}ka, V. Hovhannisyan, 
Solid State Commun. {\bf 194}, 48 (2014).

\bibitem{morita}
K. Morita, M. Fujihala, H. Koorikawa, T. Sugimoto, S. Sota, S. Mitsuda, and T. Tohyama,
Phys. Rev. B {\bf 95}, 184412 (2017)

\bibitem{ueno}
Y. Ueno, T. Zenda, Y. Tachibana, K. Okamoto and T. Sakai,
JPS Conf. Proc. {\bf 30}, 011085 (2020)


\bibitem{filho}
R. R. Montenegro-Filho, F. S. Matias, and M. D. Coutinho-Filho
Phys. Rev. B {\bf 102}, 035137 (2020)

\bibitem{hida}
K. Hida, J. Phys. Soc. Jpn. {\bf 63}, 2359 (1994). 

\bibitem{okamoto-ssc}
K. Okamoto,
Solid State Commun. {\bf 98}, 245 (1996).

\bibitem{oka-kita}
K. Okamoto and A. Kitazawa,
J. Phys. A: Math. Gen. {\bf 32}, 4601 (1999).

\bibitem{gong2}
S.-S. Gong, B. Gu, and G. Su,
Phys. Lett. A {\bf 372}, 2322(2008).

\bibitem{liu2}
G.-H. Liu, W. Li, W.-L. You, G. Su, and G.-S. Tian,
J. Mag. Mag. Mat. {\bf 377}, 12 (2015).

\bibitem{gong}
S.-S. Gong and G. Su
Phys. Rev. B {\bf 78}, 104416 (2008).

\bibitem{mahdavifar}
S. Mahdavifar and J. Abouie,
J. Phys.: Condens. Matter 23 246002 (2011).

\bibitem{jiang}
J.-J. Jiang, Y.-J. Liu, F. Tang, C.-H. Yang, Y.-B. Sheng,
Commun. Theor. Phys. {\bf 61}, 1 (2014).

\bibitem{liu3}
X.-Y. Deng, J.-Y. Dou, G.-H. Liu,
J. Mag. Mag. Mat. {\bf 392}, 56 (2015).

\bibitem{sugimoto3}
T. Sugimoto, M. Mori, T. Tohyama, and S. Maekawa
Phys. Rev. B {\bf 97}, 144424 (2018).

\bibitem{sugimoto1}
T. Sugimoto, M. Mori, T. Tohyama, and S. Maekawa,
Phys. Rev. B {\bf 92}, 125114 (2015).


\bibitem{sugimoto2}
T. Sugimoto, M. Mori, T. Tohyama, and S. Maekawa,
Phys. Rev. B {\bf 97}, 144424 (2018).

\bibitem{sasaki}
K. Sasaki, T. Sugimoto, T. Tohyama, and S. Sota,
Phys. Rev. B {\bf 101}, 144407 (2020).


\bibitem{rahaman}
Sk. S. Rahaman, M. Kumar, and S. Sahoo,
arXiv:2304.12266v2.

\bibitem{cabra}
D. C. Cabra, A. Honecker and P. Pujol, Phys. Rev. Lett. {\bf 79}, 5126 (1997). 

\bibitem{okamoto-tube}
K. Okamoto, M. Sato, K. Okunishi, T. Sakai, and C. Itoi,
Physica E {\bf 43}, 769 (2011).


\bibitem{li}
R.-. Li, S.-L. Wang, Y. Ni, K.-L. Yao, and H.-H. Fu,
Phys. Lett. A {\bf 378}, 970 (2014).


\bibitem{alecio}
Raphael C.A\'cio, M. L. Lyra, and J. Stre\v{c}ka,
J. Mag. Mag. Mat. {\bf 417}, 294 (2016).

\bibitem{farchakh}
A. Farchakh, A. Boubekri, and M. El. Hafidi, 
J. Low Temp. Phys. {\bf 206}, 131 (2022).



\bibitem{yin}
L. Yin, Z. W. Ouyang, J. F. Wang, X. Y. Yue, R. Chen, Z. Z. He, Z. X. Wang, Z. C. Xia, and Y. Liu
Phys. Rev. B {\bf 99}, 134434 (2019).

\bibitem{dey}
D. Dey, S. Das, M. Kumar, and S. Ramasesha,
Phys. Rev. B {\bf 101}, 195110 (2020).



\bibitem{yamamoto}
S. Yamamoto and T. Sakai, Phys. Rev. B {\bf 62}, 3795 (2000). 

\bibitem{sakai2}
T. Sakai and K. Okamoto, Phys. Rev. B {\bf 65}, 214403 (2002). 

\bibitem{tonegawa}
T. Tonegawa, T. Sakai, K. Okamoto and M. Kaburagi, 
J. Phys. Soc. Jpn. {\bf 76}, 124701 (2007). 

\bibitem{tenorio}
A. S. F. Ten\'orio, R. R. Montenegro-Filho, and M. D. Coutinho-Filho,
J. Phys.: Condens. Matter {\bf 23}. 506003 (2011).

\bibitem{liu}
G.-H. Liu, L.-J. Kong, J.-Y. Dou,
Solid State Commun. {\bf 213-214}, 10 (2015).


\bibitem{karlova}
K. Karlov\'a, J. Stre\v{c}ka,
Physica B {\bf 536}, 494 (2018).


\bibitem{yamaguchi}
H. Yamaguchi, T. Okita, Y. Iwasaki, Y. Kono, N. Uemoto, Y. Hosokoshi, T. Kida, T. Kawakami,
A. Matsuo,and M. Hagiwara, 
Sci. Rep. {\bf 10}, 9193 (2020).

\bibitem{cabra3}
D. C. Cabra, A. Honecker, and P. Pujol,
Phys. Rev. Lett. {\bf 79}, 5126 (1997).

\bibitem{cabra2}
D. C. Cabra, A. De Martino, A. Honecker, P. Pujol, and P. Simon,
Phys. Lett. A {\bf 268}, 418 (2000).

\bibitem{chen}
W. Chen, K. Hida and B. C. Sanctuary,
Phys. Rev. B {\bf 63}, 134427 (2001).

\bibitem{narumi}
Y. Narumi, K. Kindo, M. Hagiwara, H. Nakano, 
A. Kawaguchi, K. Okunishi, and M. Kohno,
Phys. Rev. B. {\bf 69}, 174405 (2004).

\bibitem{maximova}
for a review of $S=1$ chain compounds, O. V. Maximova, S. V. Streltsov, and A. N. Vasiliev,
Critical Reviews in Solid State and Materials Sciences, 46:4, 371 (2021).

\bibitem{yan}
X. Yan, W. Li, Y. Zhao, S.-J. Ran,  G. Su,
Phys. Rev. B {\bf 85}, 134425 (2012).

\bibitem{totsuka}
K. Totsuka, Phys. Rev. B {\bf 57}, 3454 (1998). 

\bibitem{okunishi1}
K. Okunishi and T. Tonegawa, J. Phsy. Soc. Jpn. {\bf 72}, 479 (2003). 

\bibitem{okunishi2}
K. Okunishi and T. Tonegawa, Phys. Rev. B {\bf 68}, 224422 (2003). 

\bibitem{metavitsiadis}
A. Metavitsiadis, C. Psaroudaki, and W. Brenig,
Phys. Rev. B {\bf 101}, 235143 (2020).


\bibitem{nakano}
H. Nakano and M. Takahashi, J. Phys. Soc. Jpn. {\bf 67}, 1126 (1998). 

\bibitem{okazaki1}
N. Okazaki, J. Miyoshi and T. Sakai, J. Phys. Soc. Jpn. {\bf 69}, 37 (2000). 

\bibitem{okazaki2}
N. Okazaki, K. Okamoto and T. Sakai, J. Phys. Soc. Jpn. {\bf 69}, 2419 (2000). 

\bibitem{nakasu}
A. Nakasu, K. Totsuka, Y. Hasegawa, K. Okamoto and T. Sakai, 
J. PHys.: Condens. Matter {\bf 13}, 7421 (2001). 

\bibitem{sakai3}
T. Sakai and Y. Hasegawa, Phys. Rev. B  {\bf 60}, 48 (1999). 


\bibitem{okamoto2}
K. Okamoto, N. Okazaki and T. Sakai, J. Phys. Soc. Jpn. {\bf 70}, 636 (2001). 

\bibitem{okamoto3}
K. Okamoto, N. Okazaki and T. Sakai, J. Phsy. Soc. Jpn. {\bf 71}, 196 (2002). 

\bibitem{michaud}
F. Michaud, T. Coletta, S. R. Manmana, J.-D. Picon, and F. Mila,
Phys. Rev. B {\bf 81}, 014407 (2010).

\bibitem{kohshiro}
H. Kohshiro, R. Kaneko, S. Morita, H. Katsura, and N. Kawashima
Phys. Rev. B {\bf 104}, 214409 (2021),
and refereces therein.


\bibitem{yamada}
T. Yamada, R. Nakanishi, R. Furuchi, H. Nakano, H. Kaneyasu, K. Okamoto, T. Tonegawa and T. Sakai, 
JPS Conf. Proc. {\bf 38}, 011163 (2023). 

\bibitem{PRG}
M. P. Nightingale:,
Physica A {\bf 83}, 561 (1976).

\bibitem{oka-nom}
K. Okamoto and K. Nomura,
Phys. Lett. A {\bf 169}, 433 (1992).

\bibitem{nom-oka}
K. Nomura and K. Okamoto,
J. Phys. A: Math. Gen. {\bf 27}, 5773 (1994).

\bibitem{berezinskii}
Z. L. Berezinskii: Sov. Phys. JETP {\bf 34}, 610 (1971).

\bibitem{KT}
J. M. Kosterlitz and D. J. Thouless: J. Phys. C {\bf 6}, 1181 (1973).

\bibitem{jose}
J. V. Jos\'{e}, {\it 40 Years of Berezinskii-Kosterlitz-Thouless Thoery} (World Scientific, 2013).

\bibitem{liu-ssc}
G.-H. Liu, W. Li, W.-L. You c, G. Su, and G.-S. Tian,
Solid State Commun. {\bf 166}, 38 (2013).


\bibitem{maeshima}
N. Maeshima, K. Okunishi, K. Okamoto and T. Sakai, Phys. Rev. Lett. {\bf 93}, 127203 (2004). 

\bibitem{cardy}
J. L. Cardy, J. Phys. A: Math. Gen. {\bf 17}, L385 (1984).


\end{thebibliography}

\end{document}